\documentclass[twocolumn]{aastex631}
\usepackage{breqn}

\shorttitle{Interplay Between Anisotropy and Skewness}
\shortauthors{Zenteno-Quinteros et al.}

\graphicspath{{./}{figures/}}

\begin{document}

\title{Interplay Between Anisotropy- and Skewness-driven Whistler Instabilities \\ in the Solar Wind under the Core-Strahlo model}

\author[0000-0003-2430-6058]{Bea Zenteno-Quinteros}
\affiliation{Departamento de F\'isica, Facultad de Ciencias, Universidad de Chile, Santiago, Chile} \email{beatriz.zenteno@ug.uchile.cl}

\author[0000-0002-9161-0888]{Pablo S. Moya}
\affiliation{Departamento de F\'isica, Facultad de Ciencias, Universidad de Chile, Santiago, Chile} \email{pablo.moya@uchile.cl}

\author[0000-0002-8508-5466]{Marian Lazar}
\affiliation{Centre for mathematical Plasma Astrophysics, KU Leuven, Celestijnenlaan 200B, B-3001 Leuven, Belgium}
\affiliation{Institute for Theoretical Physics IV, Faculty for Physics and Astronomy, Ruhr University Bochum, D-44780 Bochum, Germany}

\author[0000-0001-5912-5703]{Adolfo F. Vi\~nas}
\affiliation{Department of Physics \& the Institute for Astrophysics and Computational Sciences (IACS), Catholic University of America, Washington-DC, 20064, USA}  
\affiliation{NASA Goddard Space Flight Center, Heliospheric Science Division, Geospace Physics Laboratory, Mail Code 673, Greenbelt, MD 20771, USA}

\author[0000-0002-1743-0651]{Stefaan Poedts}
\affiliation{Centre for mathematical Plasma Astrophysics, KU Leuven, Celestijnenlaan 200B, B-3001 Leuven, Belgium}  
\affiliation{Institute of Physics, University of Maria Curie-Skłodowska, ul.\ Radziszewskiego 10, 20-031 Lublin, Poland}

\begin{abstract}
Temperature anisotropy and field-aligned skewness are commonly observed non-thermal features in electron velocity distributions in the solar wind. These characteristics can act as a source of free energy to destabilize different electromagnetic wave modes, which may alter the plasma state through wave-particle interactions. Previous theoretical studies have mainly focused on analyzing these non-thermal features and self-generated instabilities individually. However, to obtain a more accurate and realistic understanding of kinetic processes in the solar wind, it is necessary to examine the interplay between these two energy sources. By means of linear kinetic theory, in this paper we investigate the excitation of the parallel-propagating whistler mode, when it is destabilized by electron populations exhibiting both temperature anisotropy and field-aligned strahl or skewness. To describe the solar wind electrons, we adopt the Core-Strahlo model as an alternative approach. This model offers the advantage of representing the suprathermal features of halo and strahl electrons, using a single skew-Kappa distribution already known as the strahlo population. Our findings show that when the electron strahlo exhibits an intrinsic temperature anisotropy, this suprathermal population becomes a stronger and more efficient source of free energy for destabilizing the whistler mode. This suggests a greater involvement of the anisotropic strahlo in processes conditioned by wave-particle interactions. Present results also suggest that the contribution of core anisotropy can be safely disregarded when assessing the importance of instabilities driven by the suprathermal population. This allows for a focused study, particularly regarding the regulation of electron heat flux in the solar wind.

\end{abstract}


\section{Introduction} 
\label{sec:intro}


It is widely acknowledged that solar wind plasma can be found in nonthermal states, especially due to the high-energy, suprathermal tails (up to a few keV) of the velocity distributions of electrons and ions (protons) \citep{Collier-etal-1996, pierrard2001core, maksimovic2005radial, stverak2008electron, stverak2009radial, wilson2019aelectron, wilson2019belectron}. 
This energization is a consequence of a certain level of wave turbulence and fluctuations in the plasma system, which is not counterbalanced by the low collisionality of suprathermal particles \citep{maksimovic2005radial, wilson2019aelectron}. 
Other processes become important for the dynamics in the collisionless regime, as the free energy present in these out-of-equilibrium states is able to trigger electromagnetic or electrostatic instabilities in the plasma. 
The waves and fluctuations thus produced can interact with the charged particles, modifying the state of the plasma. 
Under this context, the most widely accepted interpretation of the solar wind electron population and its suprathermal nature is by means of three subpopulations, core, halo, and strahl, an approach that is supported by many observational reports \citep{feldman1975solar, pilipp1987variations, maksimovic1997ulysses, nieves2008solar, stverak2008electron, pierrard2016electron, tao2016quiet, wilson2019aelectron}. 
Typically measured at low energies, a dense quasi-thermal core is clearly observed in the electron distribution, a component that is usually well described by bi-Maxwellian distributions \citep{stverak2008electron, wilson2019aelectron}. 
At higher energies, we find the halo electrons enhancing the power-law tails of the distributions, which are usually modeled by bi-Kappa distributions \citep{stverak2008electron, wilson2019aelectron}. there is also the strahl population, which is a magnetic field-aligned beam, more prominent in fast winds and closer to the Sun \citep{maksimovic2005radial, stverak2009radial, bercic2019scattering, macneil2020radial, owen2022solar}. 
The strahl gives the electron distribution its distinctive skewness, a characteristic that is usually emulated in theoretical models considering a (minor) drifting anti-sunward distribution. 

The skewness provided by the strahl gives free energy to the system to eventually excite the so-called heat flux instabilities, sometimes called beaming instabilities. They receive this name as the skewness gives the electron distribution a non-zero macroscopic heat flux moment. These skewness-driven instabilities have been greatly discussed in the literature, as they are believed to be responsible for the non-collisional self-regulation of the electron heat flux in the solar wind below the collisional limit \citep{gary1977solar, gary1994whistler, gary2000whistler, bale2013electron, saeed2016electron, shaaban2018clarifying, kuzichev2019nonlinear, shaaban2019quasi, lee2019nonlinear, lopez2020alternative, micera2020particle}. In this context, recently, a new framework for studying these heat flux instabilities was proposed by \citet{zenteno2021skew}. 
These authors present a new heuristic way to model the solar wind electron distribution, called the Core-Strahlo (CS) model. The strahlo is the most important attribute of the proposed model, offering a unified description for the energetic suprathermal tails and skewness observed in the solar wind electron distributions. 
This goal has been achieved by introducing the Skew-Kappa function as a new distribution model for the electron populations. Accordingly, the CS model uses a Skew-Kappa distribution to approach the so-called "strahlo", which incorporates the suprathermal features of both the halo and strahl in a single population. Therefore, this unified description can give an alternative way of modeling solar wind electrons, which is useful for theoretical studies of instabilities, as it reduces the parameter space to be analyzed. 
Furthermore, such a combination of halo and strahl in what we now call the strahlo component in the CS model, can help to understand the interaction between these two electron populations during the expansion of the solar wind throughout the heliosphere. 
Because the halo is believed to be formed by pitch angle scattering of strahl electrons by the self-generated instabilities, e.g., heat flux instabilities \citep{maksimovic2005radial, vocks2005electron, stverak2009radial, horaites2018electron}

In \citet{zenteno2021skew} and \citet{zenteno2022role}, the authors studied in depth the whistler heat flux instability (WHFI) using the CS model. 
They showed the dependence of the whistler mode not only on the skewness parameter $\delta$, but also on other relevant plasma parameters like the strahlo beta parameter and core-to-strahlo density ratio. 
These studies were carried out considering isotropic electron populations, with skewness as the only free energy source for the generation of instabilities.
However, observations in the solar wind show that the temperature ($T$) of the electron populations is not necessarily isotropic \citep{feldman1975solar, stverak2008electron, adrian2016solar, pierrard2016electron, lazar2020characteristics}. The temperature anisotropy, $A= T_\perp /T_\parallel \ne 1$ (where $\parallel, \perp$ are directions relative to the local magnetic field), represents another source of free energy for the excitation of electromagnetic radiation, and it can be associated not only with the core but also the suprathermal populations. Instabilities triggered by temperature anisotropy have also been the subject of extensive investigation \citep{gary1996whistler, gary2006linear, camporeale2008electron, lazar2014interplay, vinas2015electromagnetic, sarfraz2016macroscopic, lazar2018temperature, lazar2018electromagnetic, shaaban2019firehose, moya2020toward, husidic2020linear, sarfraz2022electron}. For anisotropic electrons with $A>1$ the dispersion and stability theories predict two instabilities, mirror and whistler-cyclotron instability (WCI), while electrons with opposite anisotropy $A<1$ may trigger firehose instabilities, periodic and aperiodic.

As temperature anisotropies also play an important role in kinetic processes, they are an extra factor to be considered when analyzing the effect of heat flux instabilities on the dynamics of the electron population. Indeed, previous works have shown that such instabilities of, e.g., whistler mode, can cumulate the excitation effects of the temperature anisotropy and the asymmetry (or skewness) of electron populations \citep{saeed2017characteristics, shaaban2018beaming, shaaban2020whistler, sarfraz2020combined, sun2020electron}. 
Accordingly, to provide an accurate description of such kinetic effects occurring in the solar wind, it is necessary a theoretical description that takes into account the interplay between both sources of instabilities, instead of analysing them separately. 
In this work, we will focus on studying this interplay by using a new, extended CS model, which combines both sources of free energy, namely, the skewness and the intrinsic temperature anisotropies of the strahlo or core populations. 

This work is organized as follows: In Section \ref{sec:cs_model} we provide a brief summary of the new CS model, the linear kinetic theory of parallel-propagating modes, and the set of plasma parameters used in the analysis. In Section \ref{sec:strahlo} we use the extended CS model to describe the solar wind electrons and study the excitation of the whistler instability cumulating the effects of both the skewness and temperature anisotropy of the strahlo population. In Section \ref{sec:core}, we repeat the analysis but this time considering an anisotropic core instead. Finally, in Section \ref{sec:conclusions} we give a summary and present the main conclusions of this study.

\section{Core-Strahlo Model and Dispersion Relations}
\label{sec:cs_model}

The Core-Strahlo (CS) model we will use to describe the electron distribution $f_e$ consists of the superposition of two subpopulations: a quasi-thermal core and a suprathermal strahlo, as shown in Eq. (\ref{eq_total_dist}). In this framework, the core population is modeled using a drifting Bi-Maxwellian $f_c$, and a Skew-Kappa distribution function $f_s$ is used for the strahlo. The asymmetric function Eq. (\ref{eq_strahlo_dist}) is able to reproduce two important kinetic features observed in the solar wind: the enhanced high energy tails and the magnetic field-aligned skewness \citep{zenteno2021skew}. Thus, it is able to describe the distinctive features of halo and strahl subpopulations in an integrated manner.
   \begin{equation}
       \label{eq_total_dist}
        f_e(v_{\perp}, v_{\parallel}) = f_c(v_{\perp}, v_{\parallel}) + f_s(v_{\perp}, v_{\parallel})
    \end{equation}
    with: 
    \begin{equation}
         \label{eq_core_dist}
            f_c(v_{\perp}, v_{\parallel}) = \frac{n_c}{\pi^{3/2}\alpha_{\perp}^2\alpha_{\parallel}} \ \exp\left(-\frac{v_{\perp}^2}{\alpha_{\perp}^2} - \frac{(v_{\parallel}-U_{c})^2}{\alpha_{\parallel}^2}\right)    
        \end{equation}
        \begin{dmath}		
        \label{eq_strahlo_dist}
            f_{s}(v_{\perp}, v_{\parallel}) = n_sC_{s}\left[1+ \frac{1}{\kappa - \frac{3}{2}}\left(\frac{v_{\bot}^2}{\theta_{\bot}^2} + \frac{v_{\parallel}^2}{\theta_{\parallel}^2} + \delta\left( \frac{v_{\parallel}}{\theta_{\parallel}} - \frac{v_{\parallel}^3}{3\theta_{\parallel}^3}\right)\right)\right]^{-(\kappa+1)}.
        \end{dmath}	

        \begin{equation}
        \label{eq_cte_Cs}
            C_s = \dfrac{\Gamma(\kappa+1)}{\left[(\kappa - \frac{3}{2})\pi\right]^{3/2}\theta_{\perp}^2\theta_{\parallel}\Gamma(\kappa - \frac{1}{2})} \left[1-\frac{\delta^2}{4}\Upsilon_1(\kappa)\right]
        \end{equation}
        
        and:
        \begin{equation*}
            \Upsilon_1(\kappa)=\left(\dfrac{2\kappa - 1}{2\kappa - 3}\right)- \dfrac{7}{12}\,.
        \end{equation*}
        
In Eq (\ref{eq_core_dist}), the parameters $\alpha_\perp$ and $\alpha_\parallel$ are the core thermal velocities, $n_c$ is the core number density, and $U_c$ is the core drift velocity. Also, in Eq. (\ref{eq_strahlo_dist}), $n_s$ is the strahlo number density, $\delta$ is the parameter controlling the skewness of the distribution, $\kappa$ controls the slope of the high energy tails and $C_s$ is the normalization constant, such that $n_s = \int f_sd^3v$, whose expression is given in (Eq. \ref{eq_cte_Cs}). Further, $\theta_\perp$ and $\theta_\parallel$ are related to the strahlo kinetic temperatures $T_{\perp,\parallel s}$. The expression linking these parameters is $\theta_{\perp,\parallel}^2 = 2k_BT_{\perp,\parallel s}/m_e$ for the symmetric case ($\delta=0$). However, when considering skew configurations ($\delta \neq 0$) the relation gets more complicated as also involves $\delta$ and $\kappa$ (see Appendix A in \citet{zenteno2021skew} for more details). Additionally, in all expressions, the sub-indexes $\parallel$ and $\perp$ denote the parallel and perpendicular directions with respect to the background magnetic field $\mathbf{B_0} = B_0\hat{z}$.

All the mathematical details and validity range of this phenomenological model have been discussed in length in \citet{zenteno2021skew}. Simply put, for $f_s$ to be suitable as a distribution function for the solar wind electrons, we must restrict its usage to small skewness, i.e., $\delta^3 \ll 1$. Also, we must assume quasi-neutrality and zero net current, conditions that allow us to restrict the parameter space and are given by Eqs. (\ref{eq_quasineutrality}) and (\ref{eq_current_free}) in the validity range of the model. 

   \begin{equation}
    n_e = n_{c}\ + \ n_s  
    \label{eq_quasineutrality}
    \end{equation}
    \begin{equation}
        \label{eq_current_free}
         U_c = \frac{n_s}{n_c}\frac{\delta}{4}\theta_\parallel
    \end{equation}
Accordingly, in this work we use the CS distribution (\ref{eq_total_dist}) as a heuristic way to describe the solar wind electrons and study the excitation of the whistler mode due to the interplay between the skewness parameter $\delta$ and the anisotropy of both electron subpopulations, namely $A_s = \left(\frac{\theta_{\perp}}{\theta_{\parallel}}\right)^2$ and $A_c = \left(\frac{\alpha_{\perp}}{\alpha_{\parallel}}\right)^2$. The procedure to obtain the dispersive properties of electromagnetic wave modes using kinetic theory is a well-known technique. It involves the linearization of the Vlasov-Maxwell system and assuming first-order perturbations are plane waves. With this method, we are able to obtain the dispersion relation $\omega = \omega(k)$ of wave modes propagating parallel to the background magnetic field ($\mathbf{k} = k\hat{z}$) in a collisionless plasma. This is accomplished by solving the condition $|\mathcal{D}(\omega, k, f_j)| = 0$, where the dispersion tensor $\mathcal{D}$ is a 3x3 complex matrix, depending on the complex wave frequency $\omega = \omega_r + i\gamma$, the wavenumber $k$, and the distribution function of all plasma species $f_j$. Considering the mathematical limitations of the Skew-Kappa function when used to describe the electron distribution, the restriction $\delta^3 \ll 1$ allows us to obtain an analytical expression for the dispersion tensor, keeping all of the relevant physical properties of this function in the calculations. 
These expressions can be found in \citet{zenteno2021skew}, Appendix B. 

To conduct the stability analysis of the parallel propagating whistler mode, here we solve numerically the dispersion relation $\omega = \omega(k)$, applying the CS model for electrons and a Maxwellian distribution to describe the proton population ($j = p$). For this latter species, we set $\beta_{\parallel p} = 0.01$ where beta is the ratio between the kinetic and magnetic pressures. In this way, only the electrons provide free energy to the system, allowing us to study the effect anisotropy and skewness have on the whistler heat flux and whistler-cyclotron instabilities. We consider solar-wind inspired parameters throughout the study. We set the strahlo-to-core parallel temperature ratio and the kappa parameter to $T_{\parallel s}/ T_{\parallel c} = 7.0$ and $\kappa = 3.0$, respectively \citep{pierrard2016electron, lazar2020characteristics}. The strahlo relative density is set to 5\% or 10\% ($\eta_s = 0.05, 0.1$). Further, throughout this work, the skewness parameter varies between $\delta = 0.0$ and $\delta=0.25$, the anisotropies vary between $0.8 \leq A_s \leq 3.0$ for the suprathermal population and between $ 0.5 \leq  A_c \leq  3.0$ for the core population. Additionally, the magnetic field is such that the electron frequencies ratio is fixed to $\omega_{pe}/|\Omega_e| = 200$ and the beta parameter for the strahlo is set to either $\beta_{\parallel} = 0.05$ or $\beta_{\parallel} = 0.1$. These values of $\beta_{\parallel s}$ are also typically measured in the solar wind for the suprathermal population; however, they are on the lower end \citep{lazar2020characteristics}. Considering that the WCI depends strongly on beta, we chose these values to be able to easily and effectively study the interplay between the anisotropy and skewness of the electron distribution.  
 

\section{Interplay between skewness and strahlo anisotropy}
\label{sec:strahlo}

We start our linear dispersion analysis by studying the excitation of the parallel propagating whistler mode, driven unstable by a skew electron distribution with an anisotropic strahlo subpopulation. For the core, we fixed the anisotropy to $A_c = 1.0$, so that the instabilities are driven solely by the skewness and the anisotropy of the strahlo, non-thermal features controlled by $\delta$ and $A_s$, respectively. Considering that the WCI has never been studied in the context of the Core-Strahlo model, our first step is to study this instability in the symmetric case, i.e., for $\delta = 0$, so that the free energy is only provided by the strahlo anisotropy. We deem this case a reference point to later analyze the modifications introduced by the skewness. 

\begin{figure*}
\includegraphics[scale=0.22]{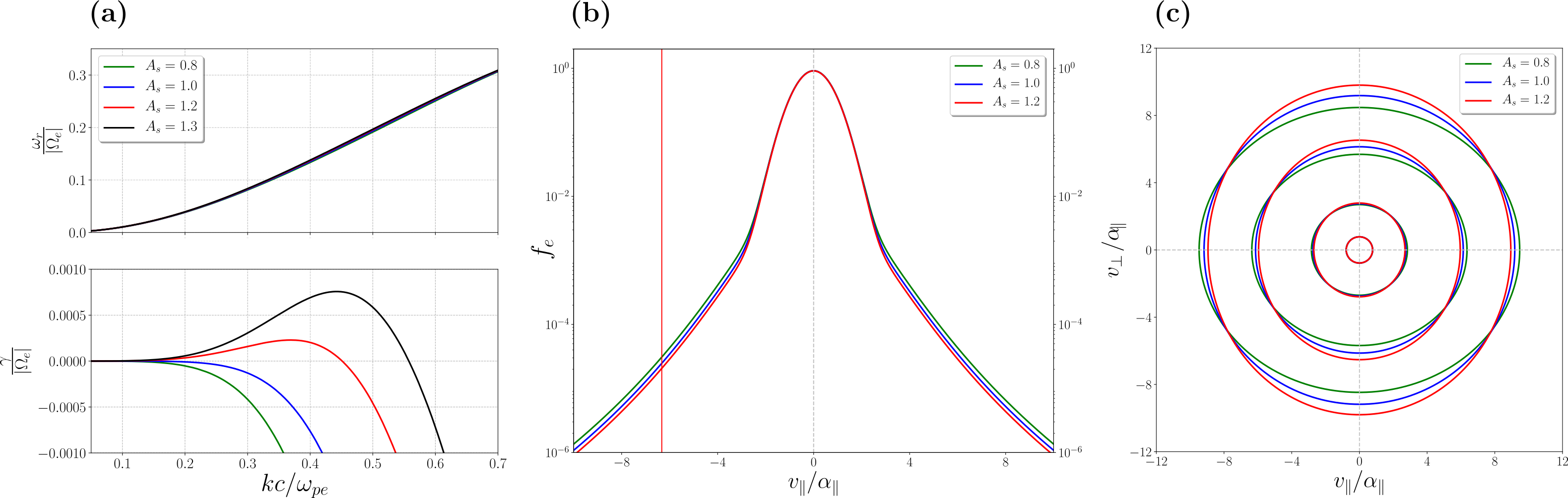}
\caption{a) Whistler mode dispersion relation for $\eta_s = 0.1$, $\beta_{\parallel s} = 0.1$, $\delta=0.0$, $A_c = 1.0$ and different values of the strahlo anisotropy $A_s$. b) Parallel cuts of the Core-Strahlo distribution function. c) Contours of the eVDF. In all panels colors represent different values of $A_s$; namely, $A_s$ = 0.8 (green), 1.0 (blue), 1.2 (red), and 1.3 (black).}
  \label{Fig_1}
\end{figure*}

In Figure \ref{Fig_1} we can see the well-known WCI. Panel \ref{Fig_1}a) shows the dispersion relation of the whistler mode for 4 different values of the strahlo anisotropy ($A_s$ = 0.8, 1.0, 1.2, and 1.3). In the top and bottom plots, we can see the real and imaginary parts of the frequency, $\omega_r$ and $\gamma$ respectively, both as a function of the wave number $k$. Frequencies are shown in units of the electron gyrofrequency $|\Omega_e|$ and wavenumbers are expressed in units of the electron inertial length $c/\omega_{pe}$. To obtain these plots, we fixed the strahlo density to 10\% and the strahlo beta parameter to $\beta_{\parallel s} = 0.1$. We can see that the real part of the frequency remains essentially the same as we modify the value of $A_s$. The imaginary part, however, depends more strongly on this parameter. The whistler mode becomes unstable ($\gamma > 0 $) only for $A_s >1$. As expected, the trend is that the higher the value of the strahlo anisotropy, the more unstable the mode becomes. The maximum growth rate $\gamma_m$ increases, and the wavenumber range in which the mode is unstable widens as $A_s$ increases. Further, for the cases shown here, we have positive growth rates for $kc/\omega_{pe} < 0.6$, which is the same wavenumber range in which the WHFI develops, and a narrower range compared to the core-driven WI, as we will see in the next section.


To see which electron configurations lead to unstable states for the whistler mode, in panels \ref{Fig_1}b) and \ref{Fig_1}c) we show, respectively, a parallel cut at $v_\perp = 0$ and a contour plot of the CS distribution (\ref{eq_total_dist}) for 3 different values of the strahlo anisotropy, namely, $A_s = 0.8$ (green), $A_s = 1.0$ (blue) and $A_s =1.2$ (red). The parallel and perpendicular velocity components (with respect to the background magnetic field) are expressed in units of the core parallel thermal speed $\alpha_\parallel$. We can see that, as we modify $A_s$ and not $A_c$, the inner contours in the 2D plot remain basically unchanged. The outer contours, however, get elongated for $A_s \neq 1$. Only one of these cases is able to excite the whistler mode, which corresponds to $A_s = 1.2$, where the distribution shape elongates in the $v_\perp$ direction (red lines). The same can be seen in the 1D plot, where only the tails are modified as we change $A_s$, while the core portion of the distribution remains the same. Moreover, in the 1D plot (panel b in the figure), we also show the resonant velocity of the most unstable mode given by Eq. (\ref{eq_res_vel}). For the $A_s = 1.2$ case, the most unstable wave configuration occurs at $kc/\omega_{pe} = 0.37$, corresponding to a maximum growth rate of $\gamma_m/|\Omega_e| = 2.3 \times 10^{-4}$. The respective real part of the frequency is $\omega_r/|\Omega_e| = 0.12$. Therefore, the resonant velocity ($v_{\rm{res}}$), given by
\begin{equation}
    \label{eq_res_vel}
    v_{\rm{res}} = \frac{\omega_r + \Omega_e}{k}
\end{equation}
has a value of $v_{\rm{res}} = -6.33\alpha_{\parallel}$, which lies in the strahlo part of the distribution. This seems reasonable, as the strahlo is the population providing the free energy to radiate. For the $A_s =1.2$ case, we also calculated the resonance term, $\xi$, for the most unstable mode according to 
\begin{equation}
    \label{eq_res_term}
    \xi = \frac{v_{\rm{res}}}{\theta_{\parallel}} = \frac{\omega_r + \Omega_e}{k\theta_{\parallel}}\,,
\end{equation}
which has a value of $\xi = -2.39$. Thus, we are describing a resonant interaction~\citep{Gary1993Theory}, which is expected for the WCI.

\subsection{Effect of skewness on the strahlo-driven whistler-cyclotron instability}

\begin{figure*}
\includegraphics[scale=0.4]{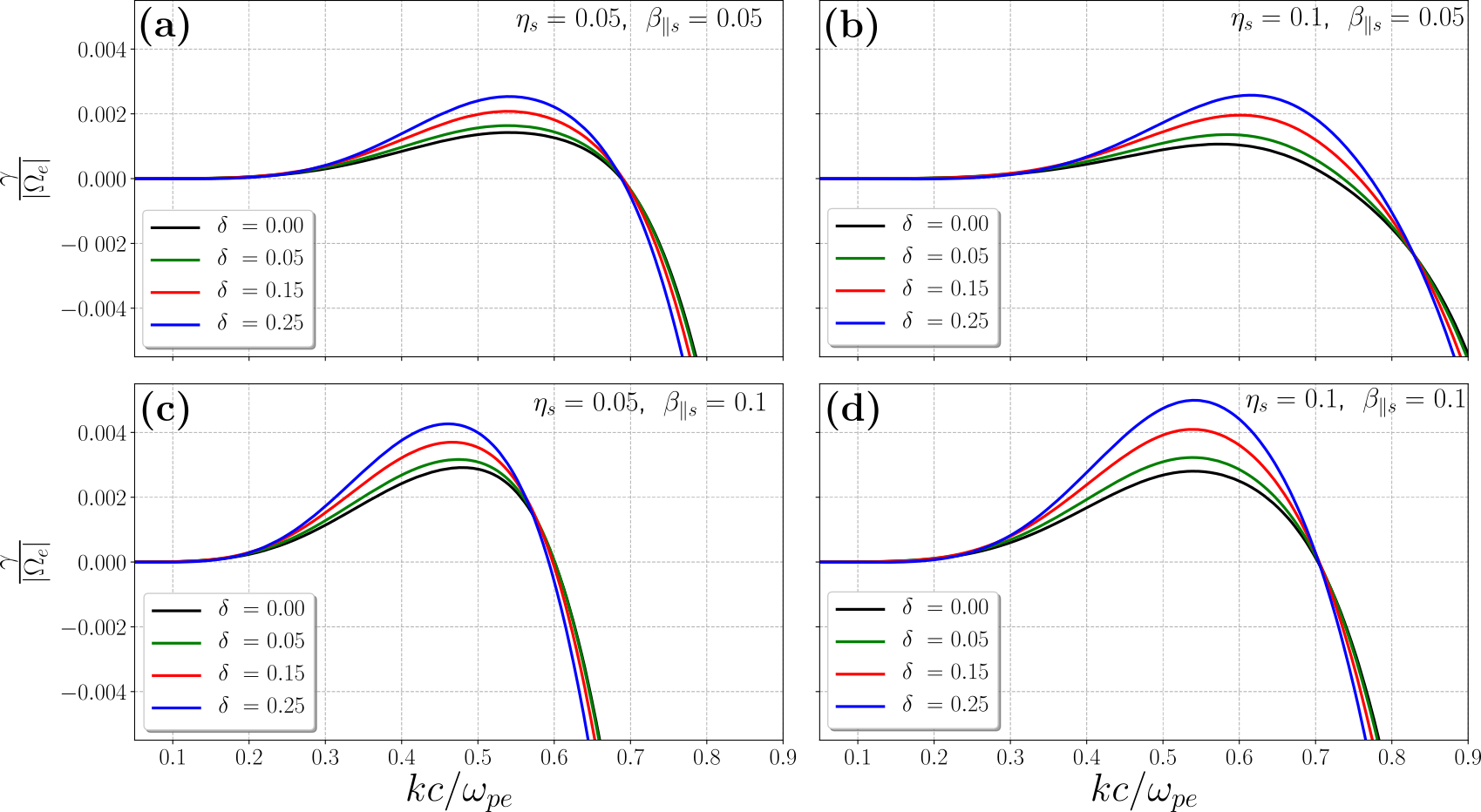}
\caption{Growth rates of the whistler mode for $A_c = 1.0$, $A_s = 1.5$ and different values of the skewness parameter: $\delta = 0.0$ (black lines), $\delta = 0.05$ (green lines), $\delta = 0.15$ (red lines) and $\delta = 0.25$ (blue lines). Also, each panel shows a different combination of strahlo density ratio and beta parameter, namely: a) $\eta_s = 0.05$ and $\beta_{\parallel s} = 0.05$, b) $\eta_s = 0.1$ and $\beta_{\parallel s} = 0.05$, c) $\eta_s = 0.05$ and $\beta_{\parallel s} = 0.1$, and d) $\eta_s = 0.1$ and $\beta_{\parallel s} = 0.1$}
  \label{Fig_2}
\end{figure*}

To see how the previous behavior is modified by the skewness parameter, we now focus our attention on cases where $\delta > 0$. Thus, we now study how the strahlo-driven WCI changes when the electron distribution also presents field-aligned skewness, which gives the system another source of free energy. Accordingly, Figure \ref{Fig_2} shows the normalized growth rates of the parallel propagating whistler mode, driven unstable by the strahlo anisotropy. For all panels, we considered an anisotropy of $A_s = 1.5$, which we know is a triggering value for the WCI and frequently measured in the solar wind for the suprathermal subpopulation. In this figure, we observe how the growth rates are modified as the skewness parameter $\delta$ increases from $\delta=0$ (black lines) to $\delta = 0.25$ (blue curves). Further, to obtain these plots, we considered different combinations of $\eta_s$ and $\beta_s$, to see how the dispersion properties also depend on these parameters. In the upper and bottom panels, we set the strahlo beta parameter to $\beta_{\parallel s} = 0.05$ and $\beta_{\parallel s} = 0.1$, respectively. Left and right panels show the growth rates for $\eta_s = 0.05$ and $\eta_s = 0.1$, respectively. Also, the respective real part of the frequencies are not shown, as they remain virtually unchanged when we modify the skewness. In all panels, we can see that the effect of $\delta >0$ is to enhance the growth rates in such a way that the higher the $\delta$ value, the more unstable the wave mode becomes. However, even though the maximum growth rate achieved increases with the skewness parameter, the range of unstable wave numbers remains essentially the same. If we now focus our attention on the left (right) panels, we can see that for a fixed strahlo relative density, increasing the value of the strahlo beta parameter enhances the growth rates, making the plasma more unstable to the whistler mode. The maximum growth rate achieved by the mode increases almost two times when we double the value of $\beta_{\parallel s}$, however, the wave-number range where the mode is unstable ($\gamma >0$) narrows with increasing beta.

Focusing on the upper panels in Figure \ref{Fig_2}, we can see that for $\beta_{\parallel s} = 0.05$, the lowest value of beta shown in the figure, the strahlo number density plays a very small role in the stability of the wave mode. We can see that when we double the value of $\eta_s$ the maximum growth rate value barely changes for the higher values of $\delta$. However, for the lowest values shown, i.e., $\delta = 0.0$ and $\delta= 0.05$, the growth rates slightly diminish. This is an unexpected behavior as we are increasing the number density of the population that gives the free energy to the waves, and maybe related to the fact that for low beta, the magnetic field dominates the dynamics of the system. Further, if we now focus on the bottom panels, we can see that for $\beta_{\parallel s} = 0.1$, we recover the expected behavior and the mode becomes more unstable with $\eta_s$ such that both the maximum growth rate and the unstable wavenumber range increase with the strahlo density ratio. In summary, the net effect of considering eVDFs with $\delta > 0$ is to enhance the strahlo-driven WCI, which was already present in the system since we are dealing with an anisotropic distribution for the suprathermal population as shown in Figure \ref{Fig_1}. This is not what happens for the core-driven WI, as we will see in the next section.

\begin{figure*}
\includegraphics[scale=0.22]{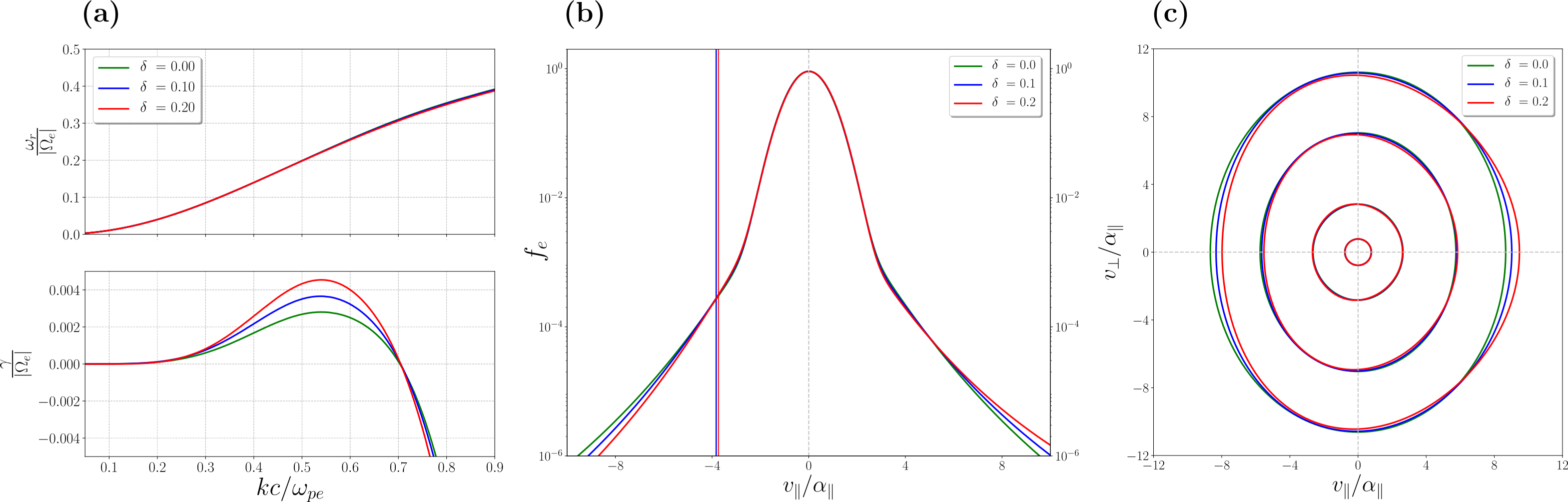}
\caption{a) Whistler mode dispersion relation for $\eta_s = 0.1$, $\beta_{\parallel s} = 0.1$, $A_s=1.5$ and $A_c = 1.0$ and different values of the skewness parameter $\delta$. b) Parallel cuts of the Core-Strahlo distribution function. c) Contours of the eVDF. In all panels colors represent different values of $\delta$; namely, $\delta$ = 0.0 (green), 0.1 (blue) and 0.2 (red)}
  \label{Fig_3}
\end{figure*}

To see how the skewness ($\delta>0$) modifies an eVDF that is already unstable to the strahlo-driven WCI, in Figure \ref{Fig_3} we show the changes that an increasing value of the skewness parameter $\delta$ introduces on the CS distribution. We also show the corresponding dispersion relations in the same fashion as in Figure \ref{Fig_1}. To obtain these plots, we considered an anisotropic electron configuration with a fixed $A_s = 1.5$, a relative strahlo density of 10\% ($\eta_s = 0.1$), and a strahlo beta parameter of $\beta_{\parallel s} = 0.1$. Accordingly, panel \ref{Fig_3}a) shows the dispersion relation for 3 different values of the skewness parameter: $\delta = 0.0$ (green lines), $\delta = 0.1$ (blue lines), and $\delta = 0.2$ (red lines). Here, we can see again that the effect of the skewness on the WCI is to make the mode more unstable, with the subsequent increase of the growth rates. Also, we can see that the real part of the frequency remains almost unchanged as we modify this parameter. Further, in panels b) and c), we show the shape of electron distributions responsible for the dispersion relations shown in \ref{Fig_3}a). Panel \ref{Fig_3}b) shows a parallel cut at $v_\perp = 0$ of the core-strahlo distribution (\ref{eq_total_dist}). Again, the parallel velocity is expressed in units of the core parallel thermal speed $\alpha_\parallel$. As expected, we can clearly see that as $\delta$ increases, the total distribution $f_e$ gets more skewed. 

Additionally, for all three unstable electron configurations, we included in this plot the resonant velocity $v$, given by Eq. (\ref{eq_res_vel}) for the most unstable mode. We can see that the resonant velocities lie, again, on the strahlo part of the distribution since that is the subpopulation providing the free energy to excite the whistler mode. Also, it is evident that the value of $v_{\rm{res}}$ does not change much as $\delta$ increases, due to the fact that both $\omega_r$ and the unstable wavenumber range seem to weakly depend on the skewness parameter. For $\delta = 0.2$, the maximum growth rate is $\gamma/|\Omega_e| = 4.5 \times 10^{-3}$, achieved at $kc/\omega_{pe} = 0.54$. The respective real frequency is $\omega_r/|\Omega_e| = 0.22$, which gives us a resonant velocity of $v_{\rm{res}}= -3.74\alpha_\parallel$ and a resonant term $\xi = -1.45$, indicating a resonant interaction. Panel \ref{Fig_3}c), on the other hand, shows a contour plot of the total electron distribution (\ref{eq_total_dist}). Here, velocities are expressed, again, in units of the core parallel thermal speed $\alpha_\parallel$. We can see that $\delta$ modifies only the outer contours, such that the distribution gets more skewed as this parameter increases. Besides, in the outer contours of this plot, we can clearly see and distinguish both non-thermal features present in the analysis (strahlo anisotropy and skewness), both contributing to the instability of the whistler mode.

\subsection{Effect of strahlo anisotropy on the whistler heat flux instability}

To continue the analysis of the combined effect the skewness parameter $\delta$ and the strahlo anisotropy $A_s$ have on the stability of the whistler mode, we now focus our attention on the effect that $A_s$ has on the WHFI, which we know is triggered by skewed electron distributions. In other words, we now study how the WHFI changes when the suprathermal population also presents another source of free energy in the form of temperature anisotropy. The WHFI was studied in detail in \citet{zenteno2021skew,zenteno2022role} for the isotropic case in the context of the CS model. The authors showed that the plasma becomes unstable to the whistler mode when $\delta > 0$, with increasing growth rates for increasing $\delta$ values. 

\begin{figure*}
\includegraphics[scale=0.4]{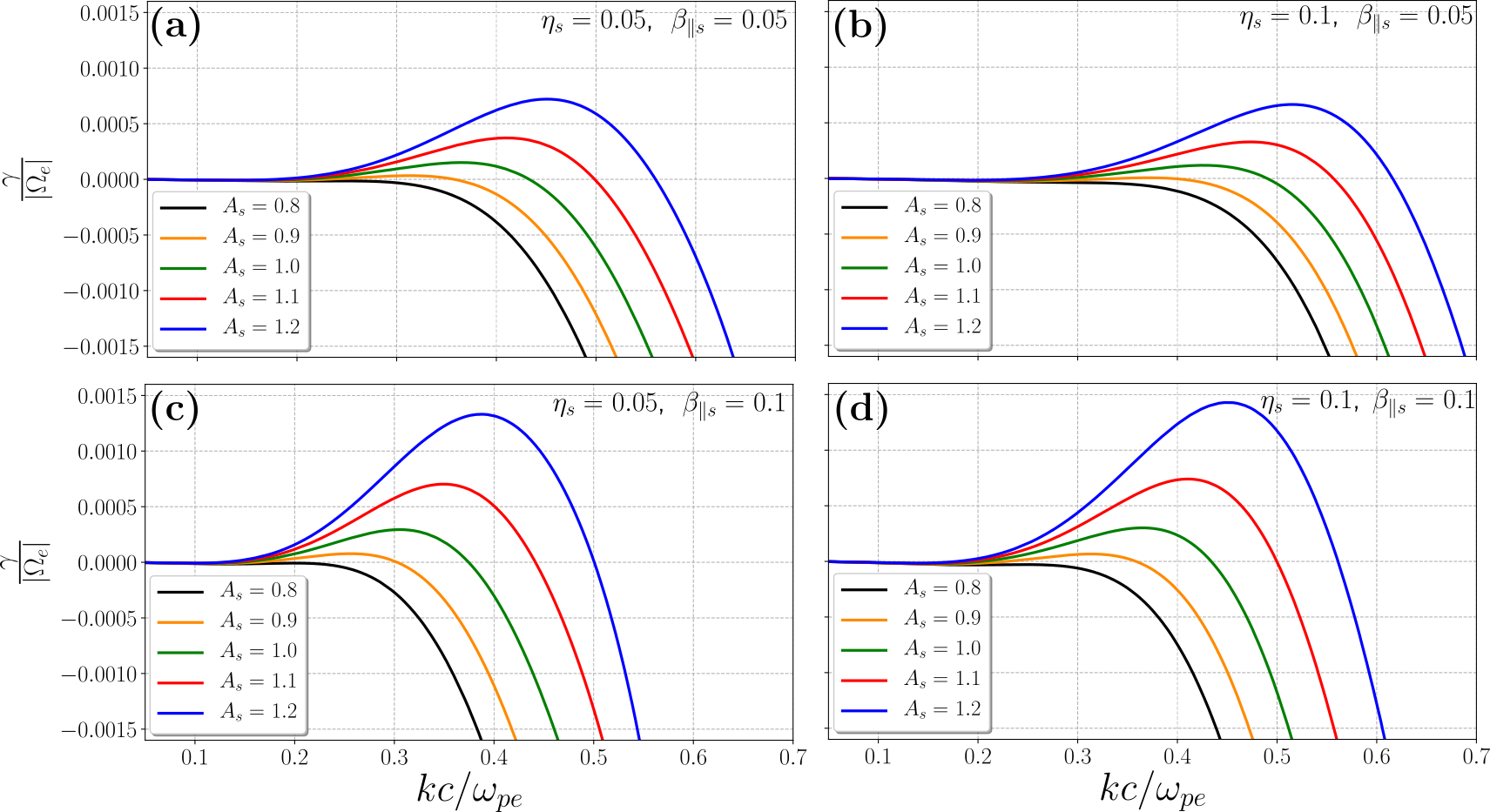}
\caption{Growth rates of the whistler mode for $A_c = 1.0$, $\delta = 0.25$ and different values of the strahlo anisotropy: $A_s = 0.8$ (black lines), $A_s = 0.9$ (orange lines), $A_s = 1.0$ (green lines), $A_s = 1.1$ (red lines) and $A_s = 1.2$ (blue lines). Also, each panel shows a different combination of strahlo density ratio and beta parameter, namely: a) $\eta_s = 0.05$ and $\beta_{\parallel s} = 0.05$, b) $\eta_s = 0.1$ and $\beta_{\parallel s} = 0.05$, c) $\eta_s = 0.05$ and $\beta_{\parallel s} = 0.1$, and d) $\eta_s = 0.1$ and $\beta_{\parallel s} = 0.1$}
  \label{Fig_4}
\end{figure*}

To see how this behavior is modified when we consider an anisotropic electron configuration, in Figure \ref{Fig_4} we show the normalized growth rates of the whistler mode, driven unstable by the strahlo skewness with $\delta =0.25$, a triggering value for the WHFI. In all panels, we can see how $\gamma$ changes as the strahlo anisotropy $A_s$ increases from $A_s = 0.8$ to $A_s = 1.2$. To obtain these plots, we considered different combinations of $\eta_s$ and $\beta_{\parallel s}$, and analyzed how the dispersion properties also depend on these plasma parameters. In the upper panels, we set $\beta_{\parallel s} = 0.05$ and in the bottom panels, we show the growth rates for $\beta_{\parallel s} = 0.1$. Further, in the left and right panels, we set the relative strahlo number density to 5\% and 10\%, respectively. The first thing we can notice in all panels is that the effect of the strahlo anisotropy is to inhibit the instability in cases where $A_s < 1$. It is known that these anisotropic cases are not able to excite the whistler mode by themselves. Thus, it seems reasonable that the same electron configurations (with $A_s < 1$) are not enhancing the WHFI. On the other hand, it is clear that plasma states with strahlo anisotropies $A_s >1$ have the opposite effect, greatly enhancing the WHFI. For these particular electron configurations, we see that as $A_s$ increases, the more unstable the whistler mode becomes. Both the maximum growth rates achieved and the unstable wavenumber range increase with $A_s$, which is the same behavior reported in \citet{shaaban2018beaming}.

Furthermore, if we focus our attention on the upper (bottom) panels and compare this dispersion property for different strahlo densities and a fixed value of $\beta_{\parallel s}$, we can see that the instability depends weakly on $\eta_s$, which is the same trend observed in Figure \ref{Fig_2}. This tells us again that the dynamics of the plasma is controlled by the magnetization at these low values of beta. For $\beta_{\parallel s}=0.05$ there is a slight decrease in the growth rate values when we double the density of the suprathermal population; however, this change is almost indistinguishable for all considered values of the anisotropy. This behavior is also observed for $\beta_{\parallel s} = 0.02$ (not shown here). Alternatively, for $\beta_{\parallel s} =0.1$ there is a slight increase in the maximum growth rate achieved, noticeable almost exclusively on the blue lines ($A_s=1.2$). Now, if we compare different values of $\beta_{\parallel s}$ for a fixed strahlo density, we can see that the mode becomes more unstable when $\beta_{\parallel s}$ increases. Maximum growth rates increase in value with beta, and the unstable wavenumber range narrows, which is the same behavior observed in Figure \ref{Fig_2}, and the expected dependence on this parameter as the plasma becomes less magnetized.

\begin{figure*}
\includegraphics[scale=0.22]{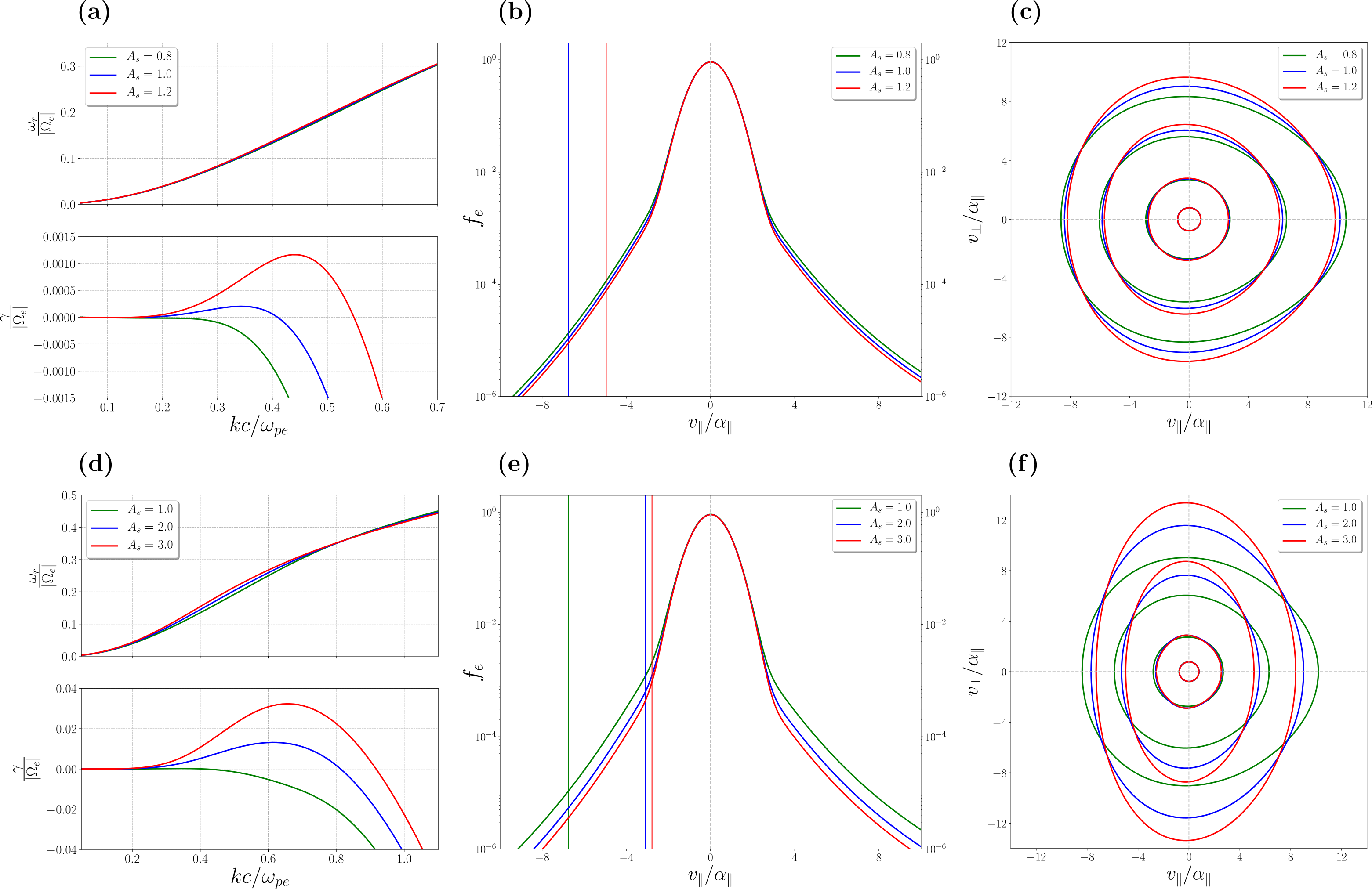}
\caption{Left panels: Whistler mode dispersion relation for $\eta_s = 0.1$, $\beta_{\parallel s} = 0.1$, $\delta=0.2$, $A_c = 1.0$ and different values of the strahlo anisotropy $A_s$. Middle panels: Parallel cuts of the Core-Strahlo distribution function. Right panels: Contours of the eVDF. In all panels colors represent different values of $A_s$; namely, $A_s$ = 0.8 (green), 1.0 (blue), 1.2 (red) (upper panels) and $A_s$ = 1.0 (green), 2.0 (blue), 3.0 (red) (lower panels)}
\label{Fig_5}
\end{figure*}

To further explore the interplay between the strahlo anisotropy and the field-aligned skewness, in Figure \ref{Fig_5} we show how changes in the value of $A_s$ modify both the dispersion relation of the parallel propagating whistler mode and the electron distribution, initially unstable to the WHFI. Accordingly, in the left panels, we can see the real and imaginary parts of the normalized frequency as a function of the normalized wavenumber for different values of the strahlo anisotropy. In the middle and right panels, we can see, respectively, parallel cuts at $v_\perp = 0$ and contour plots of the total electron distributions giving the energy to excite the whistler mode and producing the dispersion relations shown in the left panels. To obtain all of these plots, we set the strahlo relative density to 10\%, the strahlo beta parameter to $\beta_{\parallel s} = 0.1$ and $\delta = 0.2$. Upper panels show the information for moderate values of the strahlo anisotropy, more frequently measured in the solar wind \citep{lazar2020characteristics}, namely, $A_s = 0.8$ (green lines), $A_s = 1.0$ (blue lines), and $A_s = 1.2$ (red lines). From panel \ref{Fig_5}a), we can see that plasma states with $A_s > 1$ greatly enhance the instability, and states with $A_s < 1$, on the contrary, inhibit the growth rates, which is the same behavior as in the previous plot. For the particular case of $A_s = 0.8$, the instability disappears completely. From the 1D plot in panel \ref{Fig_5}b), we can see that the electron distribution is clearly skew, and the effect of changing $A_s$ on the distribution shape is to slightly enhance the tails while maintaining the core unchanged. We also include in this plot the resonant velocity $v_{res}$ of the most unstable wave mode, given by Eq. (\ref{eq_res_vel}), for the two unstable cases shown, i.e., $A_s = 1.0,\ 1.2$. We can see that the resonant velocities lie, again, on the strahlo part of the distribution, and for the latter case ($A_s = 1.2$), $v_{\rm{res}}$ is closer to 0, where there are more particles to interact with the wave. For the most unstable case $A_s=1.2$, the maximum growth rate achieved is $\gamma/|\Omega_e| = 1.2 \times 10^{-3}$ at $kc/\omega_{pe} = 0.44$, and the real part of the frequency corresponding to this $k$ value is $\omega_r/|\Omega_e| = 0.16$, which gives a resonant velocity of $v_{\rm{res}}= -4.96 \alpha_\parallel$ and a resonant term $\xi = -1.92$ meaning the interaction is resonant. From the 2D plot in panel \ref{Fig_5}c), we can see that different values of $A_s$ modify almost exclusively the outer contours while the inner ones, describing the core part of the distribution, remain almost unchanged. Further, for these moderate values of $A_s$, it is still possible to distinguish both kinetic features interacting in this situation: the field-aligned skewness and the anisotropy. 

In the bottom panels of Figure \ref{Fig_5} we show the same information as in Figure \ref{Fig_4} but this time for higher values of the strahlo anisotropy, namely, $A_s = 1.0$ (green lines), $A_s = 2.0$ (blue lines), and $A_s = 3.0$ (red lines). From panel \ref{Fig_5}d), we can see that the growth rates are greatly enhanced by the strahlo anisotropy; both the maximum growth rate and the unstable wavenumber range strongly depend on this parameter. When we double the value of $A_s$ from $A_s =1.0$ to $A_s = 2.0$, the maximum growth rate $\gamma_m$ increases almost 65 times. From the 1D plot in panel \ref{Fig_5}e), we can again distinguish the skewness of the distribution. Also, we can see that the effect of increasing $A_s$ is to diminish the energetic tails while maintaining their slope, and this time, the changes are more noticeable. For these 3 unstable configurations shown, we overplot vertical lines representing the resonant velocity $v_{\rm{res}}$ of the most unstable wave mode. We can see that as $A_s$ increases and the wave becomes more unstable, $v_{\rm{res}}$ moves to the right, even closer to 0 than in the previous case, so that there are more particles available to interact with the wave. For the most unstable case shown ($A_s=3.0$), the maximum growth rate is achieved at $kc/\omega_{pe} = 0.66$. The corresponding real and imaginary parts of the frequency are $\omega_r/|\Omega_e| = 0.29$ and $\gamma/|\Omega_e| = 3.23 \times 10^{-2}$. This gives a resonant velocity of $v/\alpha_\parallel = -2.78$ and a resonant term $\xi = -1.08$. From the 2D plot in panel \ref{Fig_5}f), we can see that changes in $A_s$ now extremely modify the outer contours of the distribution, in such a way that for $A_s=2.0,\ 3.0$ the shape is completely dominated by the strahlo anisotropy and the skewness is barely noticeable.

\section{Interplay between skewness and core anisotropy}
\label{sec:core}

We continue the linear dispersion analysis by studying the excitation of the parallel propagating whistler mode, driven unstable by an electron distribution presenting field-aligned skewness and an anisotropic core. Throughout this section, we fixed $A_s = 1.0$ to only focus on the interplay between the free energy provided by the skewness (controlled by $\delta$) and the core anisotropy (controlled by $A_c$). As in the previous Section, we start by studying the WCI for the symmetric case, i.e., $\delta = 0$, where the free energy is only provided by the anisotropy of the core ($A_c \neq 1$). We perform this analysis first because it provides a reference point to later study the modifications introduced by the skewness to the stability of the whistler mode.

\begin{figure*}
\includegraphics[scale=0.22]{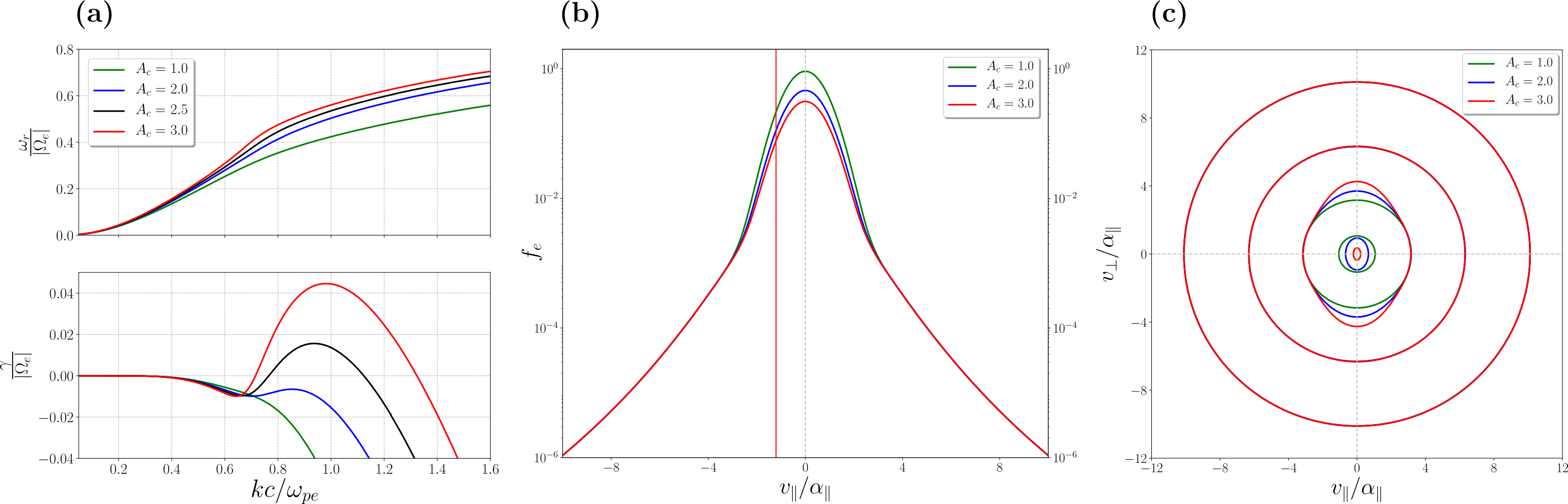}
\caption{a) Whistler mode dispersion relation for $\eta_s = 0.1$, $\beta_{\parallel s} = 0.1$, $\delta=0.0$, $A_s = 1.0$ and different values of the core anisotropy $A_c$. b) Parallel cuts of the Core-Strahlo distribution function. c) Contours of the eVDF. In all panels colors represent different values of $A_c$; namely, $A_c$ = 1.0 (green), 2.0 (blue), 2.5 (black) and  3.0 (red).}
  \label{Fig_6}
\end{figure*}

In Figure \ref{Fig_6} we show the core-driven WCI. In panel \ref{Fig_6}a) we can see the dispersion relation of the mode for different values of the core anisotropy $A_c$. The top and bottom panels show, respectively, the real and imaginary parts of the frequency, both normalized to the electron gyrofrequency $|\Omega_e|$. To obtain these plots, we fixed the strahlo density ratio to $\eta_s = 0.1$ and the strahlo beta parameter to $\beta_{\parallel s} = 0.1$. From this panel, it is clear that $\omega_r$ is more strongly affected by changes in $A_c$ when compared with other parameters, and the real frequency gets enhanced as the core anisotropy increases. Also, it is worth noticing that the values of $\omega_r$ achieved when considering an anisotropic core are higher than those achieved by the strahlo-driven WCI (see Figure \ref{Fig_1}a for comparison), which may help us differentiate the subpopulation giving the energy to excite the whistler mode. As an example, for two anisotropic configurations producing a similar maximum growth rate ($A_s = 2.0$ and $A_c = 2.4$) the respective real part of the frequency is almost two times higher when the anisotropy is provided by the core. ($\omega_r/|\Omega_e| = 0.27$ and $\omega_r/|\Omega_e| = 0.50$, respectively). Regarding the growth rates, we can see that much higher values of anisotropy are needed to excite the WCI when the energy is provided by the core population instead of the strahlo. The transition between stable and unstable modes occurs at $A_c \approx 2.2$. For lower anisotropy values, including the cases where $A_c <1$, the whistler mode is stable at all wavenumbers. This behavior, with a similar transition point, is also observed when a core-halo model with two drifting bi-Maxwellians is used to describe the electron distribution, suggesting that the presence of high-energy tails is not relevant for this instability. Further, the wavenumber range where the mode shows positive growth rates is wider for the core-driven instability when compared with the strahlo-driven one. For example, for $A_s = 3.0$ the growth rates cross the axis at $kc/\omega_{pe} \approx 0.9$ and for $A_c = 3.0$ this happens at $kc/\omega_{pe} \approx 1.35$. 

To see which electron configurations are unstable states for the whistler mode when we consider an anisotropic core, in panels \ref{Fig_6}b) and \ref{Fig_6}c) we show parallel cuts at $v_\perp = 0$ and contour plots of the CS distribution for 3 different values of the core anisotropy, namely, $A_c = 1.0$ (green), $A_c = 2.0$ (blue), and $A_c = 3.0$ (red). From the 1D plot, we can see that the effect of $A_c$ on the shape of the distribution is to lift the core portion of the distribution as $A_c$ decreases, while maintaining the high-energy tails unchanged. Here we also show the resonant velocity $v_{\rm{res}}$ of the most unstable wave mode, given by Eq. (\ref{eq_res_vel}), for the only unstable state shown, i.e., $A_c = 3.0$. For this case, the most unstable wave configuration occurs at $kc/\omega_{pe} = 0.98$, corresponding to a maximum growth rate $\gamma/|\Omega_e| = 4.5 \times 10^{-2}$ and a frequency $\omega_r/|\Omega_e| = 0.55$. This results in a resonant velocity of $v_{\rm{res}} = -1.21\, \alpha_\parallel$, which now lies in the core of the distribution. In addition, for this mode, the resonant term takes the value of $\xi = -0.46$, according to Eq. (\ref{eq_res_term}), indicating that the interaction is also resonant. On the other hand, from the 2D plot we can see that changes in $A_c$ only alter the innermost contours, which get elongated in the $v_\perp$ direction as $A_c$ increases. The outer contours, conversely, remain unchanged. This is a reasonable behavior taking into account that we are modifying a parameter particular to the core subpopulation. It is important to notice that only when the inner contours get greatly distorted and lose their oval shape, the plasma has an unstable configuration for the whistler mode.

\begin{figure*}
\includegraphics[scale=0.22]{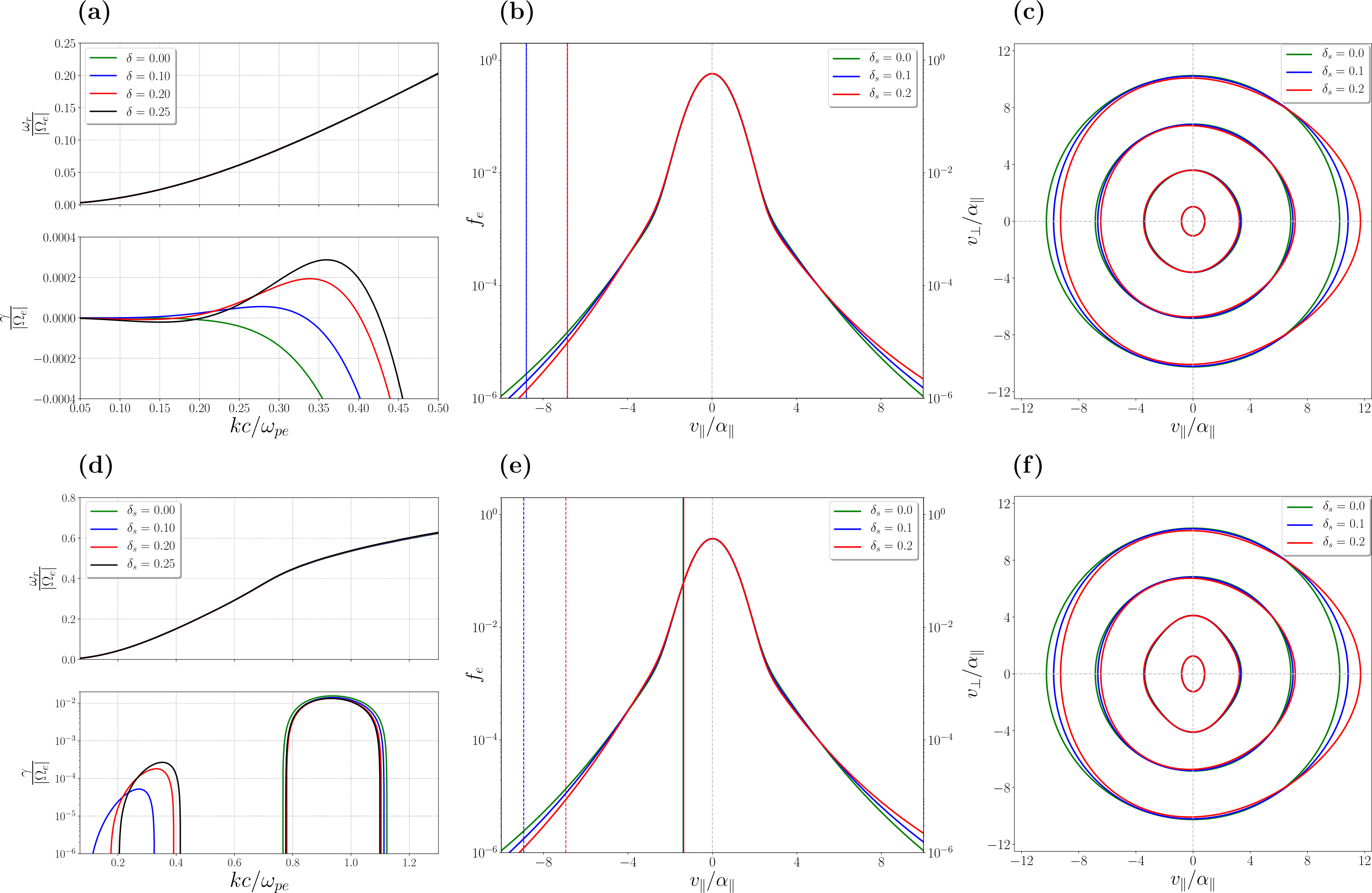}
\caption{Upper (lower) panels: Whistler mode dispersion relation, parallel cuts of the Core-Strahlo distribution function and contours of the eVDF considering $\eta_s = 0.1$, $\beta_{\parallel s} = 0.1$, $A_s = 1.0$, and $A_c = 1.6$ (top) or $A_c = 2.5$ (bottom). Different colors represent different values of the skewness parameter; namely, $\delta$ = 0.00 (green), 0.10 (blue), 0.20 (red) and 0.25 (black).}
\label{Fig_7}
\end{figure*}

\subsection{Effect of skewness on the core-driven whistler-cyclotron instability}

To see how the previous behavior changes when we include field-aligned skewness as another source of free energy, we now focus on analyzing cases where $A_c\neq 1$ and $\delta > 0$. Accordingly, in Figure \ref{Fig_7} we show the dispersion relation and the total electron distribution (\ref{eq_total_dist}) for an increasing value of the skewness parameter $\delta$, to study the modifications it introduces to the core-driven WCI. Left panels show the real and imaginary parts of the frequency, normalized to the electron gyrofrequency $|\Omega_e|$, both as a function of the normalized wavenumber and for different values of $\delta$. The middle and right panels show, respectively, parallel cuts at $v_\perp = 0$ and contour plots of the electron distributions giving the energy to excite the parallel propagating whistler mode and producing the dispersion relations shown in the left panels. To obtain all these plots, we consider a plasma state with a relative density of the suprathermal population of 10\% ($\eta_s =0.1$) and a strahlo beta parameter of $\beta_{\parallel s} = 0.1$. Further, in the upper panels, we show this information for a low value of the core anisotropy $A_c$. This means that we chose a value for the core anisotropy where, in the symmetric case ($\delta=0$), the whistler mode is always stable for any $k$ value. Considering the transition point occurs at $A_c \approx 2.2$, we chose a representative value of $A_c = 1.6$. We can see in panel \ref{Fig_7}a) that for this low value of $A_c$, the whistler mode that was initially stable (green lines) develops positive growth rates for $\delta > 0$ and gets more and more unstable as the skewness increases. Further, the wavenumbers at which $\gamma > 0$ are at least half the value of the ones typical of the core-driven WCI for the symmetric case (compare with Figure \ref{Fig_6}a). Therefore, we have a behavior distinctive of the WHFI in a wavenumber range typical of this instability driven by the skewness provided by the strahlo. Also, no onset or enhancement of the WCI driven by the core is triggered by the skewness. From panel \ref{Fig_7}b), we can see that changes in $\delta$ modify the skewness of the distribution. As expected, it becomes more skewed as $\delta$ increases while maintaining the core unchanged. We also added the resonant velocities $v_{\rm{res}}$ of the most unstable wave configuration for two unstable states: $\delta=0.1$ and $\delta = 0.2$. We can see that these velocities lie in the strahlo part of the distribution, as expected. Thus, for low values of $A_c$ the instability is completely triggered by the skewness of the strahlo. Moreover, for $\delta = 0.2$ the maximum growth rate, the corresponding wavenumber and real part of the frequency are, respectively, $\gamma/|\Omega_e| = 1.9 \times 10^{-4}$, $kc/\omega_{pe} = 0.34$, and $\omega_r/|\Omega_e| = 0.11$. This gives us a resonant velocity of $v_{\rm{res}} = -6.85\,\alpha_\parallel$ and a resonant term of $\xi = -2.66$ (a resonant interaction). Also, as the plasma gets more unstable to the mode, the resonant velocity moves to the right, to higher values of $f_e$, so that the density of particles able to interact with the wave gets higher as $\delta$ increases. Further, from the 2D plot in panel \ref{Fig_7}c), we can clearly see in the outer contours that the core-strahlo distribution gets more skew as $\delta$ increases while the inner contours remain unchanged. Also, the elongated inner contours in the $v_\perp$ direction indicate that the eVDF presents a core anisotropy $A_c>1$. However, they are not distorted enough, with respect to the isotropic case, to be able to excite the WCI.

To continue the analysis, in the bottom panels of Figure \ref{Fig_7} we show the same information regarding the effect of the skewness parameter, but this time for a high value of the core anisotropy. Here we chose a value of $A_c =2.5$ (larger than 2.2), representing an unstable plasma state for the WCI in the symmetric case ($\delta =0$). From the dispersion relations shown in panel \ref{Fig_7}d) we can see again that $\delta$ does not seem to have an important effect on the real frequency $\omega_r$ of the whistler mode. The imaginary part, on the other hand, depends more strongly on this parameter. For this particular case, we show the growth rates in log scale, so we can easily distinguish the two peaks with $\gamma>0$ that appear for the same mode. The peak present at higher $k$ values is at least two orders of magnitude more intense than the one present at lower wavenumbers. Further, the more intense peak lies in a wavenumber range typical of the core-driven WCI. Consistently, we can see that changes in the $\delta$ parameter do not seem to have a significant effect on the growth rates values for this specific $k$ range. In contrast, the secondary peak appears at a wavenumber range typical of instabilities driven by the strahlo subpopulation; the mode gets more unstable as $\delta$ increases, which is the distinctive behavior of the WHFI. Further, the peak present at higher $k$ values (associated with core-driven WCI) has corresponding real frequencies which are at least 4 times higher than the ones corresponding to the secondary peak (associated with the WHFI). For $\delta = 0.25$, the corresponding real frequencies are $\omega_r/|\Omega_e| = 0.51$ and $\omega_r/|\Omega_e| = 0.12$, respectively. This may provide a useful way to observationally distinguish the subpopulation providing the free energy to the instability.

Moreover, from the 1D plot in panel \ref{Fig_7}e) we can see the same behavior as in panel \ref{Fig_7}b), the distribution shape gets more skewed as $\delta$ increases, maintaining the core unchanged. In the plot, we also included the resonant velocities (\ref{eq_res_vel}) of the most unstable wave configuration of each peak for two selected unstable states, namely, $\delta= 0.1$ (blue) and $\delta = 0.2$ (red). Solid lines correspond to the resonant velocity of the WCI, and we can see that they lie in the core portion of the distribution and are not modified significantly by changing the skewness of the distribution. For $\delta = 0.2$, the most unstable wave configuration for the most intense peak is located at $kc/\omega_{pe} = 0.93$, with a maximum growth rate $\gamma/|\Omega_e| = 1.3 \times 10^{-2}$ and a frequency $\omega_r/|\Omega_e| = 0.51$, which gives a resonant velocity of $v_{\rm{res}} = -1.36\,\alpha_\parallel$ and a resonant term of $\xi = -0.52$. Dashed lines, on the other hand, correspond to the resonant velocities of the WHFI. We can see that they lay on the tails of the distribution corresponding to the strahlo subpopulation, and they have the same behavior as in panel \ref{Fig_7}b). Namely, the resonant velocity moves to the right as the plasma gets more unstable, so that the density of particles able to interact with the wave gets higher as $\delta$ increases. For $\delta = 0.2$, the secondary peak occurs at $kc/\omega_{pe} = 0.33$, with a maximum growth rate of $\gamma/|\Omega_e| = 1.8 \times 10^{-4}$ and a frequency of $\omega_r/|\Omega_e| = 0.11$. This gives a resonant velocity $v_{\rm{res}}= -6.95\,\alpha_\parallel$ and a resonant term $\xi = -2.7$. Again, these resonant terms show that instability is given by a resonant interaction, which also proves that the instability should not be mistaken with the firehose instability. Finally, from the 2D plot in panel \ref{Fig_7}f) we can clearly differentiate the two sources of free energy in this system. The distribution exhibits elongated and distorted inner contours, responsible for the core-driven WCI. We can see that these contours do not change considerably as $\delta$ increases, which is consistent with the fact that the stability of the mode remains almost unchanged for variations of $\delta$. In the outer contours, we can clearly see the skewness of the CS distribution, which gives the energy for the less intense WHFI. Therefore, we recover the typical behavior of the WHFI: as $\delta$ increases, the distribution gets more skewed and the plasma gets more unstable, as the second peak shows in panel \ref{Fig_7}d).  

\begin{figure*}
\includegraphics[scale=0.22]{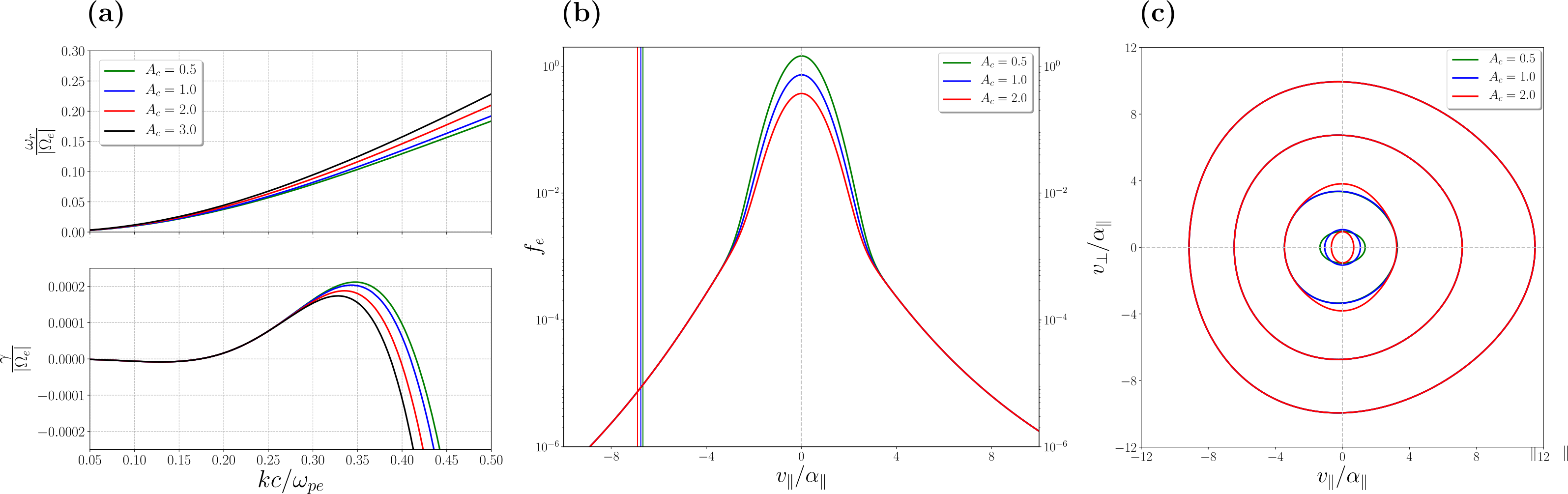}
\caption{a) Whistler mode dispersion relation for $\eta_s = 0.1$, $\beta_{\parallel s} = 0.1$, $\delta=0.2$, $A_s = 1.0$ and different values of the core anisotropy $A_c$. b) Parallel cuts of the Core-Strahlo distribution function. c) Contours of the eVDF. In all panels colors represent different values of $A_c$; namely, $A_c$ = 0.5  (green), 1.0 (blue), 2.0 (red), and 3.0 (black).}
  \label{Fig_8}
\end{figure*}


\subsection{Effect of core anisotropy on the whistler heat flux instability}

To finalize the study of the interplay between the skewness parameter $\delta$ and the core anisotropy $A_c$, we now focus on the effect that $A_c$ has on the WHFI, which we know is triggered by a skewed electron distribution ($\delta \neq 0$). i.e., we now study how the WHFI modifies in the presence of an anisotropic core distribution, giving another source of free energy to the system. We know that, for the isotropic case $A_c=1$, the plasma becomes unstable when $\delta > 0$ and as this parameter increases, the electron distribution becomes more skewed and the plasma becomes more unstable to the parallel propagating whistler mode. To see how this behavior is modified when we consider an anisotropic core subpopulation, in Figure \ref{Fig_8} we show how changes in $A_c$ modify both the dispersion relation of the whistler mode and the electron distribution initially unstable to the WHFI. To obtain all these plots, we set the strahlo density to 10\%, the strahlo beta parameter $\beta_{\parallel s} = 0.1$, and we consider a skewed distribution with $\delta=0.2$. The left panel shows the real and imaginary parts of the normalized frequency as a function of the normalized wavenumber for different values of the core anisotropy $A_c$. Here we can see that $\omega_r$ depends more strongly on $A_c$ compared with $\delta$, and the frequencies increase as the core anisotropy increases. Regarding the growth rates, we can see that this time cases with $A_c < 1$ enhanced the instability. Both the maximum growth rate value achieved and the unstable wavenumber range where $\gamma >0$ increase for anisotropy values less than one. On the other hand, cases with $A_c > 1$ have the opposite effect and diminish the instability as $A_c$ increases. Both the maximum growth rate value and the unstable wavenumber range diminish as the core anisotropy increases. Note that nonetheless, the changes introduced by the core anisotropy are very small, especially considering the high values of $A_c$ shown in this plot. Thus, the WHFI does not depend as strongly on the core anisotropy when compared with $A_s$, and we can say that anisotropic core populations do not have a significant effect on enhancing or diminishing the WHFI. Also, we are not showing higher wavenumber $k$ values, where the WCI appears, as we already know how it depends on $A_c$, and also know from figure \ref{Fig_7} that $\delta$ does not have any significant effect on its stability. 

To explore this interplay further, we look again at the electron distribution functions triggering the excitation of the whistler mode and how they change as we modify $A_c$. Accordingly, the middle and right panels show, respectively, parallel cuts at $v_\perp = 0$ and contour plots of the total electron distributions giving the energy to excite the whistler mode and producing the dispersion relations shown in the left panels. The green, blue, and red curves represent, respectively, $A_c = 0.5,\ 1.0$, and $2.0$. From the 1D plot shown in panel b), we can see that the effect of decreasing the core anisotropy is to raise the core portion of the distribution, while maintaining the tails unchanged. We also plot here the resonant velocity of the most unstable wave configuration for the three unstable cases shown. We can see the velocities lie in the strahlo part of the distribution and almost do not vary as $A_c$ increases. For $A_c = 2.0$, the maximum growth rate, the corresponding wavenumber and real part of the frequency are, respectively, $\gamma/|\Omega_e| = 1.8 \times 10^{-4}$, $kc/\omega_{pe} = 0.33$, and $\omega_r/|\Omega_e| = 0.11$. This gives a resonant velocity of $v_{\rm{res}} = -6.89\,\alpha_\parallel$ and a resonant term of $\xi = -2.67$. Furthermore, from the 2D plots of the core-strahlo distributions shown in panel c) we can see that, as expected, changes in $A_c$ only modify the innermost contours we identify with the core subpopulation, and the outer contours remain unchanged, which is consistent with the small effect $A_c$ has on the WHFI.

The effect of $A_c$ on the whistler mode is the opposite behavior as the one reported in \citet{shaaban2018beaming} (see Figure 2). To obtain this figure, the authors used a Core-Halo model for electron population, where each component was described by a drifting bi-Maxwellian according to Eq. (\ref{eq_core_dist}), so that the skewness of the eVDF is provided exclusively by the relative drift velocity, $U_h$, between Core and Halo. Moreover, in their analysis, the value of the relative drift they selected was such that the electron distribution presented two distinct peaks. This configuration is not achievable by the CS model and represents another source of free energy inter-playing in the system, in addition to the skewness and anisotropy. Moreover, if we use this Core-Halo description to study the effect of the core anisotropy $A_c$ on the WHFI, we can recover the behavior shown in Figure \ref{Fig_8}a) by considering a low value for the drift between Core and Halo. By low value of the drift, we mean a differential streaming that can provide a skewness, but the peak due to the halo is not visible in the electron distribution. For this configuration, cases with $A_c <1$ enhance the instability of the whistler mode, and cases with $A_c >1$ diminish it, which further proves the applicability of the CS model as a useful description of the solar wind electron population. However, it should be mentioned that in the Core-Halo description with drifting bi-Maxwellians, these low values of the drift are only able to trigger the WHFI for large enough values of the beta parameter. This is because the whistler mode does not get unstable for all values of the drift satisfying $U_h > 0$, but rather there is a threshold that must be exceeded to excite the WHFI. Besides, this threshold value for the drift is highly dependent on beta. For example, for the values of beta we have been using throughout this work for the suprathermal population, namely, $\beta_{\parallel h} = 0.05,\ 0.1$ the threshold value is $U_h/\alpha_{\parallel h} = 0.55,\ 0.3$, respectively, where $\alpha_{\parallel h}$ is the thermal velocity of the Halo subpopulation. On the contrary, in the CS description, the WHFI gets triggered for all values of $\delta >0$, even for low beta values. This suggests that for the WHFI, the enhanced high energy tails present in the CS description are an important feature to be considered and, thus, should be included for a realistic assessment of the instability in the solar wind.

\section{Summary and Conclusions}
\label{sec:conclusions}

We have used linear kinetic theory to perform a systematic study on the excitation of the parallel propagating whistler mode, triggered by electron populations presenting skewness and anisotropy, in a collisionless and magnetized plasma. To study the interplay between these suprathermal features and their combined effect on the stability of the whistler mode, we used the Core-Strahlo model~\citep{zenteno2021skew} for the electron population. This model uses a superposition of two subpopulations to describe the electron distribution: a quasithermal core and a suprathermal strahlo (main carrier of the heat flux under this representation), described by a Bi-Maxwellian and a Skew-Kappa distribution, respectively. In the applicability range ($\delta^3 \ll 1$), the CS model is able to reproduce the three main kinetic characteristics of the solar wind electrons, namely, the quasithermal core, enhanced high-energy tails, and field-aligned skewness. Here we consider an extended CS model, which combines two sources of free energy (skewness and temperature anisotropy). It is important to note that the CS model as a theoretical tool is limited to small skewness ($\delta \ll 1$). The model was proposed with the intention of simplifying the description of the electrons and the subsequent kinetic studies of their dynamics, but it has not been observationally tested yet.

However, we believe the model has interesting properties that should be explored further. Besides, several other models have been proposed with alternative ways of describing the suprathermality of the electron distribution, which, to our knowledge, have also not been observationally tested \citep{scherer2018regularized, horaites2018stability, vasko2019whistler, husidic2020linear}. This shows there is an interest to extend and diversify the study of kinetic processes, as the use of these kinds of heuristic models can still give us important theoretical insights regarding the solar wind dynamics to be later tested with observations. Additionally, most works where stability analyses are performed do not provide plots of the shape of the unstable distribution functions, so we do not see how they change with different values of relevant plasma parameters. Considering the distributions are key to knowing when a mode will become unstable for given plasma conditions, understanding how different suprathermal features affect the excitation of a given instability (and to which extent they modify the mode's stability) may be important to gain intuition about the system, regardless of the theoretical model chosen to describe the system. Thus, throughout the linear stability analysis of whistler mode we performed in this article, we have systematically shown the corresponding electron distributions. 

To isolate the effect of each subpopulation, we first considered a skewed electron distribution with an anisotropic strahlo. We studied the stability of the whistler mode and its sensitivity to the parameters controlling these features, namely, $A_s$ and $\delta$. We later repeated the analysis but considering an isotropic but skewed strahlo and an anisotropic core, and studied the changes on the dispersive properties introduced by the interplay between $A_c$ and $\delta$. For both cases, we solved the dispersion relation $\omega=\omega(k)$ of the whistler mode numerically, considering typical solar wind conditions for the rest of the relevant plasma parameters. Additionally, throughout the study, we showed the total electron distribution for a representative number of unstable configurations and showed how the shape of the distribution changes for different values of the anisotropies and skewness. Namely, $A_s$, $A_c$ and $\delta$. In this manner, we were able to understand to which extent each parameter impacts the excitation and stability of the whistler mode. Further, this allows us to have a visual representation of the suprathermal features operating in the system and providing the energy to radiate waves. Thus, regardless of the mathematical model chosen to represent solar wind electrons, through the analysis here presented, we gained intuition about which non-thermal feature is more relevant in exciting the whistler mode. 

Our results regarding the combined effect of $A_s$ and $\delta$ on the linear stability of the whistler mode showed that both of the strahlo-driven instabilities, namely, the WCI and the WHFI, manifest themselves as positive growth rates in similar wavenumber ranges when compared with the core-driven WCI. Thus, making it hard to differentiate the feature (temperature anisotropy or skewness) of the suprathermal population providing the energy to the system. However, for reasonable values of $A_s>1$ typically measured in the solar wind, the strahlo anisotropy is able to produce higher growth rates and in a broader wavenumber range when compared with the instabilities driven by skewness. As an example, the WCI triggered by a plasma state with $A_s = 1.25$ and $\delta=0$ produces a maximum growth rate of $\gamma/|\Omega_e| = 4.49 \times 10^{-4}$, which is 2.2 times higher than the one produced by the WHFI triggered by a plasma state with a strong skewness of $\delta=0.2$ and $A_s=1$. Further, for these values of $A_s > 1$, typically measured in the solar wind, the changes introduced by $\delta$ on the strahlo-driven WCI are not significantly strong and are negligible when considering higher values of $A_s$. When we consider an anisotropic plasma with $A_s = 1.5$, the maximum growth rate of the instability increases more than 1.5 times when we go from the symmetric case $\delta=0$, to a strong value of skewness $\delta = 0.2$. 

On the contrary, the effect of the strahlo anisotropy $A_s$ on the WHFI is stronger. For a skewed plasma configuration with $\delta=0.2$, the maximum growth rate of the instability increases 2.8 times when we go from the isotropic case $A_s = 1.0$ to a plasma configuration with a slightly higher anisotropy of $A_s =1.1$. These results suggest that the anisotropy of the suprathermal population is a more effective and potent source of free energy to destabilize the whistler mode in comparison to the field-aligned skewness. Therefore, considering that anisotropy produces more explosive instabilities, this linear analysis suggests that $A_s$ is a more important parameter to focus on when studying wave-particle interaction processes and the non-collisional relaxation of the electron population to quasi-stable states in the solar wind. However, the energy exchange between particles and electromagnetic fields is a complex and highly non-linear process. Indeed, some studies have shown that more explosive instabilities (with higher growth rates) may saturate very fast, which may not allow the necessary time to have an effective energy exchange. Conversely, instabilities producing smaller linear growth rates may be sustained for longer, thus having stronger effects over time \citep{lopez2019particle}. Therefore, a quasilinear and/or non-linear approach is necessary to fully understand the interplay between anisotropy and skewness and the relative role the respective instabilities have on the solar wind electron dynamics. 

On the other hand, the WCI triggered by the core population manifests itself as positive growth rates at a larger wavenumbers when compared with strahlo-driven instabilities (triggered by the anisotropy or skewness of the strahlo), which may allow us to differentiate the subpopulation providing the free energy to the system. Along the same lines, the real part of the frequency, $\omega_r$, of the strahlo-driven instabilities are consistently lower than the ones produced by the core-driven WCI, which may represent another way to distinguish the energy source from observations. Further, much more extreme values of $A_c$ are needed to excite the whistler mode when compared to $A_s$. The transition between always stable and unstable occurs at $A_c \approx 2.2$ when the energy is provided by the core. Conversely, when the energy is provided by the strahlo, all cases with $A_s>1$ are unstable. After that, similar values of anisotropy produce similar growth rates for both cases. For example, a plasma system with an anisotropic strahlo (core) with $A_s = 2.5$ ($A_c = 2.5$) produces a maximum growth rate of $\gamma/|\Omega_e| = 1.92 \times 10^{-2}$ ($\gamma/|\Omega_e| = 1.56 \times 10^{-2}$). Additionally, the strahlo skewness (controlled by $\delta$) does not have a particularly strong effect on the core-driven WI. As we increase this parameter, both the unstable wavenumber range and maximum growth rates achieved by the instability stay almost the same. Reciprocally, the changes introduced by the core anisotropy $A_c \neq 1$ on the growth rates of the WHFI are also small. When we consider a skewed distribution with $\delta=0.2$ the maximum growth rate achieved increases 1.08 times when we cut in half the value of the core anisotropy from $A_c=2.0$ to $A_c=1.0$. These results suggest that we can safely ignore the contribution of the core anisotropy when assessing the importance of different instabilities driven by the suprathermal population on the solar wind electron dynamics. This allows to reduce the parameter space to be studied, which is especially pertinent when we focus on the electron heat flux regulation problem and the relevance of the WHFI in this process. 

Nonetheless, it is worth noting that throughout this work, we have restricted the analysis to electron configurations with only one anisotropic subpopulation at a time. However, this restriction does not apply to solar wind electrons, where anisotropy can be present in both the core and suprathermal subpopulations. It is likely that these two sources of anisotropy to be relevant and contribute to the electron dynamics. Therefore, a more in-depth analysis, allowing both subpopulations to simultaneously exhibit anisotropy and going beyond linear theory, is necessary to fully understand the interplay between these two suprathermal characteristics, and assess their importance in the kinetic processes of the solar wind. Accordingly, we intend to employ electromagnetic PIC simulations to rigorously test and further enhance the analysis presented in this study, encompassing a broader range of realistic solar wind scenarios. However, such an extensive analysis is beyond the scope of the present investigation. 

The results here shown are consistent with previous works and with more established models for the solar wind electron population, which gives theoretical support to the CS model. This representation for the electron population allows us to reproduce the main kinetic features of the solar wind electron distributions, using a unified representation for the suprathermal population, namely, halo and strahl. This description uses fewer free parameters than the usual core-halo-strahl representations, which may allow to study more easily the interaction between these two subpopulations in intermediate states where the halo is forming at the expense of the strahl. Also, it may allow to develop simpler theoretical studies focusing on the effect that electron configurations with a not so prominent strahl have on non-collisional processes that may influence the solar wind dynamics. Therefore, we hope the results shown here will motivate the development of theoretical works studying the dynamics of the halo and strahl using a unified description. This may be helpful to deepen and extend the analysis performed in this work and include the influence of oblique propagating whistler modes in the electron distribution shape. Also, to consider the interplay with other instabilities known to exist in the solar wind, like the firehose instability. Thus, using a unified description for the halo and strahl may be especially helpful when addressing the complex dynamics of interaction between these suprathermal populations as they move away from the sun.

\subsection*{acknowledgments}
We thank the support of ANID, Chile through the  Doctoral National Scholarship N$^{\circ}$21181965 (B.Z.Q.) and FONDECyT grant No. 1191351 (P.S.M.). A. F.- Viñas would like to thank the Catholic University of America/IACS and NASA-GSFC for their support during the development of this work. The authors acknowledge support from the Ruhr-University Bochum and the Katholieke Universiteit Leuven. These results were also obtained in the framework of the projects C14/19/089 (C1 project Internal Funds KU Leuven), G.0025.23N (FWO-Vlaanderen), SIDC Data Exploitation (ESA Prodex-12), Belspo project B2/191/P1/SWiM.






\bibliography{manuscript}{}
\bibliographystyle{aasjournal}



\end{document}